\documentclass[a4paper,12pt]{article}

\usepackage{amsmath}
\usepackage[dvips]{graphicx}
\def\be{\begin{equation}}
\def\eeq{\end{equation}}
\newcommand{\en}{\end{equation}}
\def\ba{\begin{eqnarray}}
\def\ea{\end{eqnarray}}

\def\D{\nabla}

\def\U{\mathcal{U}}

\def\Dc{\mathcal{D}}
\def\oQ{\overline{Q}}
\def\oD{\overline{\nabla}}
\def\oa{\overline{a}}
\def\olambda{\overline{\lambda}}
\def\orho{\overline{\rho}}
\def\oG{\overline{G}}
\def\os{\overline{s}}

\renewcommand{\theequation}{\arabic {section}.\arabic{equation}}
\makeatletter
\@addtoreset{equation}{section}
\makeatother

\begin{document}
%%%%%%%%%%%%%%%%%%%%%%%%%%%%%%%%%%%%%%%%%%%%%%%%%%%%%%%%%%%%%%%%%%%%%%%%%%%%

\begin{flushright}
DFTT 2007/15 
\\
EPHOU 07-004 
\\
RIKEN-TH-111
\\
July, 2007
\end{flushright}

\vfil

\begin{center}

{\Large Exact Extended Supersymmetry on a Lattice: \\

\vspace{0.3cm}

Twisted $N=4$ Super Yang-Mills in Three Dimensions}\\

\vspace{1cm}

{\sc Alessandro D'Adda}\footnote{dadda@to.infn.it}$^{a}$, 
{\sc Issaku Kanamori}\footnote{kanamori-i@riken.jp}$^{b}$
{\sc Noboru Kawamoto}\footnote{kawamoto@particle.sci.hokudai.ac.jp}$^{c}$ and 
{\sc Kazuhiro Nagata}\footnote{knagata@indiana.edu}$^{d}$\\

\vspace{0.5cm}
$^{a}${\it{ INFN sezione di Torino, and Dipartimento di Fisica Teorica, Universita 
di Torino, I-10125 Torino, Italy }}\\
$^{b}${\it{ Theoretical Physics Laboratory, RIKEN }}\\
{\it{ Wako, 351-0198, Japan}}\\
$^{c}$
{\it{ Department of Physics, Hokkaido University }}\\
{\it{ Sapporo, 060-0810, Japan}}\\
and \\
$^{d}${\it{ Department of Physics, Indiana University }}\\
{\it{ Bloomington, 47405, IN, U.S.A.}}
\end{center}

\vfil

%%%%%%%%%% Abstract %%%%%%%%%%
\begin{abstract}
We propose a lattice formulation of three dimensional super Yang-Mills  
model with a twisted $N=4$ supersymmetry. The extended supersymmetry algebra 
of all the eight supercharges is fully and exactly realized on the lattice 
with a modified ``Leibniz rule". 
The formulation we employ here is a three dimensional extension of
manifestly gauge covariant method which was developed in our 
previous proposal of Dirac-K\"ahler
twisted $N=2$ super Yang-Mills on two dimensional lattice. 
The twisted $N=4$ supersymmetry algebra is geometrically realized on a 
three dimensional lattice with link supercharges and the use of
``shifted" (anti-)commutators.
A possible solution to the recent critiques on the link formulation will 
be discussed.

\end{abstract}

%%%%%%%%%%%%%%%%%%%%%%%%%%%%%%%%%%%%%%%%%%%%%%%%%%%%%%%%%%%%%%%%%%%%%%%%%%%%
%%%%%%%%%%%%%%%%%%%%%%%%%%%%%%%%%%%%%%%%%%%%%%%%%%%%%%%%%%%%%%%%%%%%%%%%%%%%

\newpage

% Reset parameters
\setcounter{footnote}{0}

\section{Introduction }

Formulating an exact supersymmetric model on a lattice is one of the most
challenging subjects in lattice field theory.
There has been already a number of works addressing this topic 
\cite{DKKN1,DKKN2,old-lattsusy,D-W-lattsusy,deconstruction,
other-lattsusy1,twist-lattsusy,%twist-lattsusy2,
other-lattsusy2,
numerical-lattsusy,Leib-fujikawa,Japanese-recent}. 
Recently, it has been recognized that 
a so-called twisted version of supersymmetry (SUSY) plays a particularly important role
in formulating supersymmetric models on a lattice 
\cite{DKKN1,DKKN2,twist-lattsusy}.
The crucial importance of twisted SUSY on the lattice
could be traced back to %the recognition of
the intrinsic relation between twisted fermions
and Dirac-K\"ahler fermion formulation \cite{KT}.
Based on this recognition, in \cite{DKKN1} we proposed lattice formulations of
super BF and Wess-Zumino models based on 
Dirac-K\"ahler twisted $N=2$ chiral and anti-chiral superfields on two dimensional lattice,
and then in \cite{DKKN2} we proceeded to formulate a manifestly gauge covariant formulation
of twisted $N=2$ super Yang-Mills (SYM) action on two dimensional lattice. 
The main feature of our formulation is that 
``Leibniz rule" on the lattice can be exactly maintained
throughout the formulation, and as a result,
the resulting lattice action is %can be shown to be
invariant w.r.t. all the supercharges 
associated with the twisted SUSY algebra.
It has been also recognized in \cite{DKKN2} that, besides twisted $N=2$ in two dimensions,
Dirac-K\"ahler twisted $N=4$ SUSY algebra in four dimension \cite{KKM}
could also be realized on the lattice with the lattice Leibniz rule.
In this paper, we point out that $N=4$ twisted SUSY
algebra in three dimensions, which has eight supercharges,
can also be consistent with the lattice Leibniz rule requirements
and then present an explicit construction of
corresponding $N=4$ SYM action on three dimensional lattice. 

In recent papers the authors of \cite{Bruckmann,BC} posed some critiques 
on our formulations of noncommutative approach \cite{DKKN1} and the link 
approach \cite{DKKN2}. A possible answer to the critique on the 
noncommutative approach\cite{DKKN1} will be given by analyses of a 
matrix formulation of super fields\cite{ADKS}. 
Along the similar line of arguments to the critique in the noncommutative 
approach, we propose a possible solution to the link approach\cite{DKKN2} 
with which we share the same treatment in this paper.

\section{Discretization of $N=4$ twisted SUSY algebra in three dimensions}

We first introduce the following $N=4$ SUSY algebra
in Euclidean three dimensional continuum spacetime. 
\begin{eqnarray}
&&\hspace{20pt}
\{Q_{\alpha i},\overline{Q}_{j\beta}\} = 2\delta_{ij}(\gamma_{\mu})_{\alpha\beta}P_{\mu}
\label{algebra}\\[2pt]
[J_{\mu},Q_{\alpha i}] &=& +\frac{1}{2}(\gamma_{\mu})_{\alpha\beta}Q_{\beta i},
\hspace{20pt}
%\\[0pt]
%[J_{\mu},\overline{Q}_{i\alpha}] &=& -\frac{1}{2}
%\overline{Q}_{i\beta}(\gamma_{\mu})_{\beta\alpha}\\[0pt]
[R_{\mu},Q_{\alpha i}]\ =\ -\frac{1}{2}Q_{\alpha j}(\gamma_{\mu})_{ji}\\[2pt]
%[R_{\mu},\overline{Q}_{i\alpha}] &=& +\frac{1}{2}
%(\gamma_{\mu})_{ij}\overline{Q}_{j\alpha}\\[2pt]
[J_{\mu},P_{\nu}] &=& -i\epsilon_{\mu\nu\rho}P_{\rho}, \hspace{13pt} %\\[4pt]
[J_{\mu},J_{\nu}] \ =\ -i\epsilon_{\mu\nu\rho}J_{\rho},\hspace{10pt} %\\[2pt]
[R_{\mu},R_{\nu}] \ =\ -i\epsilon_{\mu\nu\rho}R_{\rho},\\[4pt]
[R_{\mu},P_{\nu}] &=&
[P_{\mu},P_{\nu}] \ =\ 
[J_{\mu},R_{\nu}] \ =\ 0 %\\[2pt]
\end{eqnarray}
where 
%$\overline{Q}_{i\alpha}$ denotes the complex conjugation
%of $Q_{\alpha i}$, $\overline{Q}_{i\alpha}=Q^{*}_{\alpha i}$, and
the gamma matrices, $\gamma_{\mu}$,  
%\begin{eqnarray}
%\{\gamma^{\mu},\gamma^{\nu}\} = 2 \delta^{\mu\nu}\mathbf{1}
%\end{eqnarray}
can be taken as Pauli matrices,
$\gamma^{\mu}(\mu=1,2,3)\equiv (\sigma^{1},\sigma^{2},\sigma^{3})$.  
$J_{\mu}$ and $R_{\mu}\ (\mu =1,2,3)$ are the generators for $SO(3)$ Lorentz
and internal rotations, respectively.
$\overline{Q}_{i\alpha}$ can be taken as the complex conjugation
of $Q_{\alpha i}$, $\overline{Q}_{i\alpha}=Q^{*}_{\alpha i}
=Q^{\dagger}_{i\alpha}$ in the continuum spacetime.

As in $N=D=2$ or $N=D=4$ case \cite{KT,KKM}, twisting procedure can be performed
through introducing twisted Lorentz generator $J'_{\mu}$ as
a diagonal sum of original Lorentz and internal rotation generators,
$J'_{\mu} \equiv J_{\mu} + R_{\mu}$.
%\begin{eqnarray}
%J'_{\mu} \equiv J_{\mu} + R_{\mu},
%\end{eqnarray}
Since, after the twisting, the Lorentz indices $\alpha$ and the internal indices $i$
are treated in the equal footing,
the resulting algebra is most naturally expressed by means of the following
Dirac-K\"ahler 
expansion of the supercharges on the basis of gamma matrices,
\begin{eqnarray}
Q_{\alpha i} &=& (\mathbf{1}Q+\gamma_{\mu} Q_{\mu})_{\alpha i}, \hspace{30pt}%\\[4pt]
\overline{Q}_{i\alpha} \ =\ (\mathbf{1}\overline{Q}+\gamma_{\mu}\overline{Q}_{\mu})_{i\alpha},
\end{eqnarray}
where $\mathbf{1}$ represents two-by-two unit matrix.
The coefficients of the above expansions, $(Q,\oQ_{\mu},Q_{\mu},\oQ)$,
are called twisted supercharges of $N=4$ in three dimensional
continuum spacetime. 
%which transform as a (scalar, vector, vector, scalar), respectively,
%under the twisted Lorentz generator $J'_{\mu}$.
After the twisting and the expansions, the original SUSY algebra (\ref{algebra})
can be expressed as,
\begin{eqnarray}
\{Q,\oQ_{\mu}\} &=& P_{\mu}, \label{D=3alg1} \\[2pt]
\{Q_{\mu},\oQ_{\nu}\} &=& -i\epsilon_{\mu\nu\rho}P_{\rho}, \label{D=3alg2}\\[2pt]
\{\oQ,Q_{\mu}\} &=& P_{\mu}, \label{D=3alg3}
\end{eqnarray}
where $\epsilon_{\mu\nu\rho}$ is three dimensional totally anti-symmetric 
tensor with $\epsilon_{123}=+1$. 
One could  see that the twisted supercharges $(Q,\oQ_{\mu}, Q_{\mu},\oQ)$
transform as (scalar, vector, vector, scalar), respectively, under
the twisted Lorentz generator $J'_{\mu}$ in the continuum spacetime.
Although the above type of $N=4$ twisted SUSY algebra in three dimensions
has been discussed also in the context of topological field theory \cite{GM},
we would rather proceed, in the following,
 to formulate a possible lattice counterpart of the algebra
(\ref{D=3alg1})-(\ref{D=3alg3}).

As it was discussed in details in \cite{DKKN1,DKKN2} that one should 
maintain the Leibniz rule to realize exact SUSY
on a lattice. 
The importance of Leibniz rule has also been recognized in the context of
non-commutative differential geometry on a lattice \cite{KK}.
Let us remind some generic argument of the formulation.
Since we have only finite lattice spacings on a lattice, infinitesimal 
translations should be replaced by finite difference operators,
\begin{eqnarray}
P_{\mu}\ =\ i\partial_{\mu} \rightarrow i\Delta_{\pm\mu}
\end{eqnarray}
where $\Delta_{\pm\mu}$ denote forward and backward difference
operators, respectively.
The operation of $\Delta_{\pm\mu}$ on a function $\Phi(x)$
can be defined by the following type of ``shifted" commutators,
\begin{eqnarray}
(\Delta_{\pm\mu} \Phi(x)) &\equiv& \Delta_{\pm\mu} \Phi(x)
- \Phi(x\pm n_{\mu}) \Delta_{\pm\mu}, \label{D_Phi}
\end{eqnarray}
which satisfy the following ``lattice" Leibniz rule,
\begin{eqnarray}
(\Delta_{\pm\mu} \Phi_1(x)\Phi_2(x)) &=& (\Delta_{\pm\mu}\Phi_1(x))\Phi_2(x)
+ \Phi_1(x\pm n_{\mu}) (\Delta_{\pm\mu}\Phi_2(x)), \label{D_Phi2}
\end{eqnarray}
where the $\Delta_{\pm\mu}$, locating on links from $x$ to $x\pm n_{\mu}$,
respectively, take unit values for generic $x$,
\begin{eqnarray}
\Delta_{\pm\mu}\ =\ (\Delta_{\pm\mu})_{x\pm n_{\mu},x} \ =\ \mp 1.
\end{eqnarray}
Since the lattice formulation of SUSY should embed 
the above properties of bosonic operators into the SUSY algebra, it is natural 
to assume that a lattice SUSY transformation can also be defined
as a ``shifted" (anti-)commutator of $Q_{A}$
located on a link from $x$ to $x+a_{A}$,
\begin{eqnarray}
(Q_A\Phi(x)) &\equiv& (Q_{A})_{x + a_{A},x}\Phi(x) 
- (-1)^{|\Phi|} \Phi(x + a_{A})(Q_{A})_{x+a_{A},x}, \label{ops}
\end{eqnarray}
where $|\Phi|$ represents $0$ or $1$ for bosonic or fermionic field $\Phi$,
respectively.
The operation of $Q_{A}$'s on a product of fields
accordingly gives,
%\footnote{Refs. \cite{Bruckmann,BC} pose some critics 
%discussing the interchanging (anti-)symmetry
%between $\Phi_{1}(x)$ and $\Phi_{2}(x)$.
%We should emphasize here that the operation of $Q_{A}$ given in 
%(\ref{Q_Phi_Phi%})
%is manifestly (anti-)symmetric under interchanging $\Phi_{1}(x)$ and 
%$\Phi_{2}(x)$,
%as is clearly seen if (\ref{ops}) is substituted into (\ref{Q_Phi_Phi}),
%although some more aspects of semi-local nature associated with 
%component fields remain to be better treated in the future development.}
\begin{eqnarray}
(Q_{A}\Phi_{1}(x)\Phi_{2}(x))
&=&
(Q_{A}\Phi_{1}(x))\Phi_{2}(x) 
+ (-1)^{|\Phi_{1}|} \Phi_{1}(x+a_{A})(Q_{A}\Phi_{2}(x)). \label{Q_Phi_Phi}
\end{eqnarray}
Since the supercharges $Q_{A}$'s are located on links,
it is then natural to define  
an anti-commutator of lattice supercharges %is to be defined
as an successive connection of link operators,
\begin{eqnarray}
\{Q_{A},Q_{B}\}_{x+a_{A}+a_{B},x} =
(Q_{A})_{x+a_{A}+a_{B},x+a_{B}} (Q_{B})_{x+a_{B},x} 
+ (Q_{B})_{x+a_{A}+a_{B},x+a_{A}} (Q_{A})_{x+a_{A},x}. \label{link_com}
\nonumber \\
\end{eqnarray} 

\begin{figure}
\begin{center}
\begin{minipage}{70mm}
\begin{center}
\includegraphics[width=50mm]{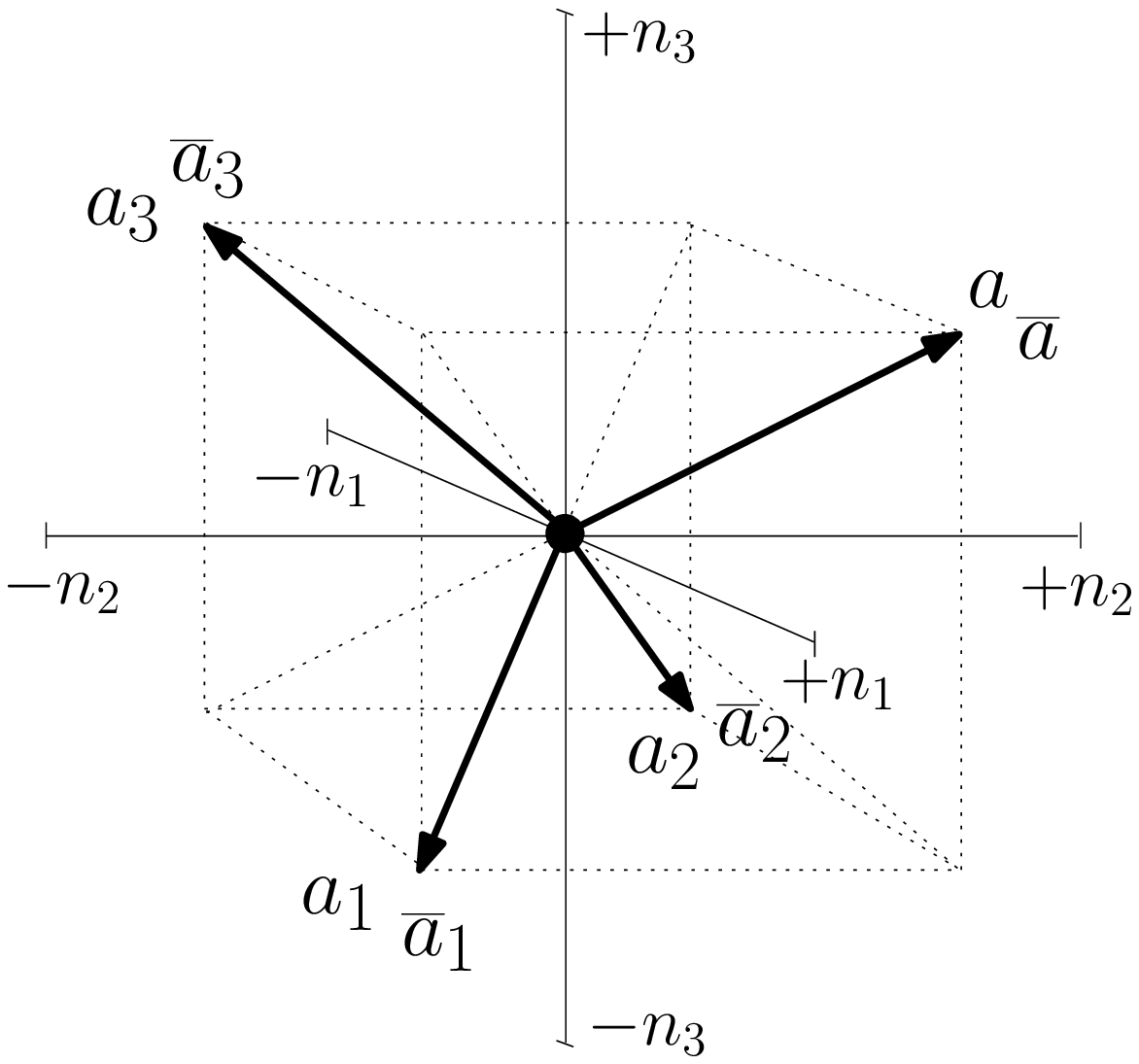}
\caption{Symmetric choice of $a_{A}$}
\label{3Dsymm_a}
\end{center}
\end{minipage}
\begin{minipage}{65mm}
\begin{center}
\includegraphics[width=70mm]{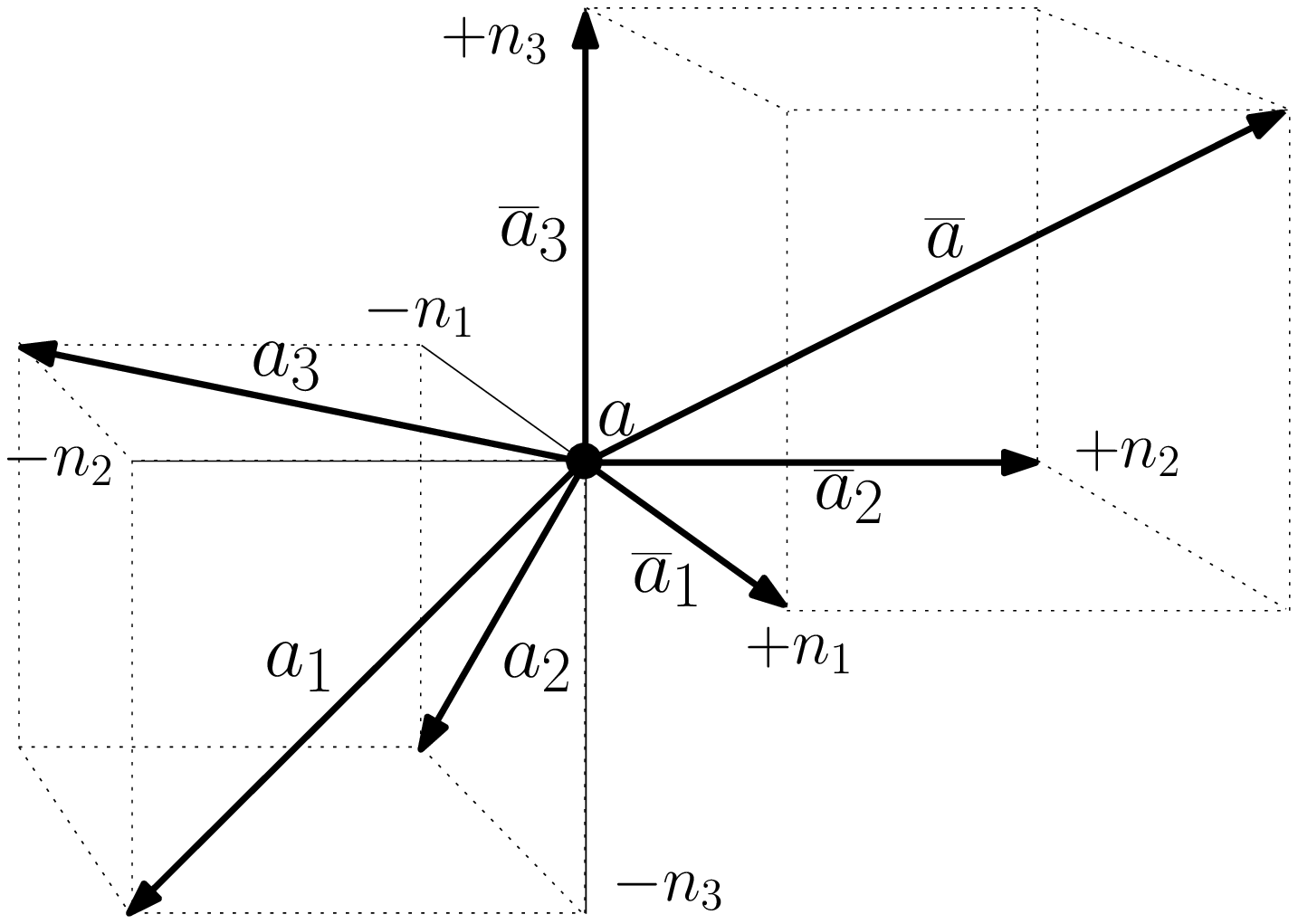}
\caption{Asymmetric choice of $a_{A}$}
\label{3Dasymm_a}
\end{center}
\end{minipage}
\end{center}
\end{figure}

By means of the above ingredients, 
lattice SUSY algebra could be expressed as
\begin{eqnarray}
\{Q_{A},Q_{B}\}_{x+a_{A}+a_{B},x} &=& 
(\Delta_{\pm\mu})_{x\pm n_{\mu},x}, \label{expected_alg} %\\[2pt]
\end{eqnarray}
provided the following lattice Leibniz rule conditions hold
\begin{eqnarray}
a_{A}+ a_{B} &=& +n_{\mu} \hspace{20pt} for\ \ \ \Delta_{+\mu},  \label{llcond1}\\
a_{A}+ a_{B} &=& -n_{\mu} \hspace{20pt} for\ \ \ \Delta_{-\mu}, \label{llcond2}
\end{eqnarray}
which are the necessary conditions for the realization of lattice SUSY algebra
and eventually govern the structure of supersymmetric lattices.
As described in \cite{DKKN1,DKKN2}, one could show that 
Dirac-K\"ahler twisted type of
$N=D=2$ and $N=D=4$ SUSY algebra can satisfy the conditions.
We point out here that the lattice counterpart of
$N=4\ D=3$ twisted SUSY algebra 
introduced in (\ref{D=3alg1})-(\ref{D=3alg3})
could also satisfy the conditions and be expressed as
\footnote{Altogether $2^{3}=8$ possible
choices of forward or backward difference operators
are consistent with the lattice Leibniz rule.},
\begin{eqnarray}
\{Q,\oQ_{\mu}\} &=& +i\Delta_{+\mu}, \label{D=3latalg1} \\[2pt]
\{Q_{\mu},\oQ_{\nu}\} &=& +\epsilon_{\mu\nu\rho}\Delta_{-\rho}, \label{D=3latalg2}\\[2pt]
\{\oQ,Q_{\mu}\} &=& +i\Delta_{+\mu}, \label{D=3latalg3}
\end{eqnarray}
where the anti-commutators of the l.h.s are understood as shifted 
anti-commutators.
The corresponding Leibniz rule conditions,
\begin{eqnarray}
a+\overline{a}_{\mu} &=& +n_{\mu}, \label{leibniz1}\\
a_{\mu} + \overline{a}_{\nu} &=& -|\epsilon_{\mu\nu\rho}|n_{\rho}, \label{leibniz2}\\
\overline{a}+a_{\mu} &=& +n_{\mu}, \label{leibniz3}
\end{eqnarray}
could be consistently satisfied by the following generic solutions,
\begin{eqnarray}
a &=& (arbitrary), \hspace{30pt}
\overline{a}_{\mu} \ =\ +n_{\mu} - a, \\
a_{\mu} &=& -\sum_{\lambda \neq \mu} n_{\lambda} +a, \hspace{30pt}
\overline{a} \ =\ +\sum_{\lambda =1}^{3} n_{\lambda} -a.
\end{eqnarray}
Notice that there is one vector arbitrariness 
in the choice of $a_{A}$, 
which governs the possible configurations of three dimensional lattice.
One of the typical examples is the symmetric choice (Fig.\ref{3Dsymm_a}) given by,
\begin{eqnarray}
a &=& (+\frac{1}{2},+\frac{1}{2},+\frac{1}{2}),\hspace{25pt}
\overline{a} \ =\  (+\frac{1}{2},+\frac{1}{2},+\frac{1}{2}), \label{3Dsymm1}\\
a_1 &=& (+\frac{1}{2},-\frac{1}{2},-\frac{1}{2}),\hspace{20pt}
\oa_1 \ =\  (+\frac{1}{2},-\frac{1}{2},-\frac{1}{2}), \label{3Dsymm2}\\
a_2 &=& (-\frac{1}{2},+\frac{1}{2},-\frac{1}{2}),\hspace{20pt}
\oa_2 \ =\  (-\frac{1}{2},+\frac{1}{2},-\frac{1}{2}), \label{3Dsymm3}\\
a_3 &=& (-\frac{1}{2},-\frac{1}{2},+\frac{1}{2}),\hspace{20pt}
\oa_3 \ =\  (-\frac{1}{2},-\frac{1}{2},+\frac{1}{2}), \label{3Dsymm4}
\end{eqnarray}
and the other one is the asymmetric choice (Fig.\ref{3Dasymm_a}) characterized by
\begin{eqnarray}
a &=& (0,0,0),\hspace{43pt}
\oa \ =\  (+1,+1,+1), \label{3Dasymm1}\\[2pt]
a_1 & =&  (0,-1,-1), \hspace{19.5pt}
\oa_1 \ =\ (+1,0,0), \label{3Dasymm2}\\[2pt]
a_2 & =&  (-1,0,-1), \hspace{19.5pt}
\oa_2 \ =\ (0,+1,0), \label{3Dasymm3}\\[2pt]
a_3 & =&  (-1,-1,0), \hspace{19.5pt}
\oa_3 \ =\ (0,0,+1).\label{3Dasymm4}
\end{eqnarray}
Notice that the summation of all the shift parameters
$(a,\overline{a}_{\mu},a_{\mu},\overline{a})$ vanish,
\begin{eqnarray}
\sum a_{A} &=& a + \overline{a}_{1} +\overline{a}_{2} + \overline{a}_{3}
+ a_{1} +a_{2} +a_{3} + \overline{a} \ =\ 0, 
\end{eqnarray}
regardless of any particular choice of $a_{A}$.
Each of the above two choices exhibits the important characteristics
of twisted $N=4\ D=3$ lattice supercharges. For asymmetric choice
(\ref{3Dasymm1})-(\ref{3Dasymm4}), one could see that
each supercharge of $(Q,\oQ_{\mu},Q_{\mu},\oQ)$ is
located on (site, link, face, cube), respectively,
which covers all the possible fundamental simplices on three dimensional 
simplicial manifold.
This observation justifies the reason why we need
eight components of supercharges to be 
embedded on three dimensional lattice.
On the other hand, for the case of symmetric choice, 
(\ref{3Dsymm1})-(\ref{3Dsymm4}),
each $a_{A}$ and $\oa_{A}$
coincide with each other, namely,
$a = \oa,\ a_{1}  = \oa_{1},\  a_{2} = \oa_{2},\ a_{3} = \oa_{3}$, 
and only four corners of %out of the eight in
the three dimensional cube 
are occupied by the link supercharges. 
This ``degenerated" structure of link supercharges
is a peculiar property to the odd dimensional lattice,
which should eventually be related to the absence of 
chirality in odd dimensions.

%According to the Dirac-K\"ahler fermion formulation \cite{....},
%three dimensional fermions could be expanded by the 
%simplicial elements of 0-form, 1-form, 2-form and 3-form,
%\begin{eqnarray}
%\Psi &=& \psi + \psi_{\mu}dx^{\mu} +\psi_{\mu\nu}dx^{\mu}\wedge dx^{\nu}
%+\psi_{\mu\nu\rho}dx^{\mu}\wedge dx^{\nu}\wedge dx^{\rho}
%\end{eqnarray}

\section{Lattice formulation of twisted $N=4$
SYM in three dimensions}

Based on the arguments in the previous section
we now proceed to construct $N=4$ twisted 
SYM action on Euclidean three dimensional lattice
along the similar manner as in $N=D=2$ twisted lattice
SYM \cite{DKKN2}.
We first introduce fermionic and bosonic gauge link variables,
$\D_{A}$ and $\U_{\pm\mu}$ which are located on links
$(x+a_{A},x)$ and $(x\pm n_{\mu},x)$, respectively,
just like $Q_{A}$ and $\Delta_{\pm \mu}$.
The gauge transformations of those link operators
are given by,
\begin{eqnarray}
(\D_{A})_{x+a_{A},x} &\rightarrow& G_{x+a_{A}}(\D_{A})_{x+a_{A},x}G^{-1}_{x}, \\
(\U_{\pm \mu})_{x\pm n_{\mu},x} &\rightarrow& 
G_{x\pm n_{\mu}}(\U_{\pm \mu})_{x\pm n_{\mu},x}G^{-1}_{x},
\end{eqnarray}
where $G_{x}$ denotes the finite gauge transformation at the site $x$. 
Next we impose the following $N=4$ twisted
SYM constraints on three dimensional lattice,
\begin{eqnarray}
\{\D,\overline{\D}_{\mu}\}_{x+a+\overline{a}_{\mu},x}
&=& +i(\U_{+\mu})_{x+n_{\mu},x}, \label{constSYM1}\\
\{\D_{\mu},\overline{\D}_{\nu}\}_{x+a_{\mu}+\overline{a}_{\nu},x}
 &=& -\epsilon_{\mu\nu\rho}(\U_{-\rho})_{x-n_{\rho},x},\label{constSYM2}\\
\{\overline{\D},\D_{\mu}\}_{x+\overline{a}+a_{\mu},x} 
&=& +i(\U_{+\mu})_{x+n_{\mu},x}, \label{constSYM3}\\
\{others\} &=& 0, \label{constSYM4}
\end{eqnarray}
where the left-hand sides should be
understood as link anti-commutators such as (\ref{link_com}), for example,
\begin{eqnarray}
\{\D,\oD_{\mu}\}_{x+a+\oa_{\mu},x}
&=& (\D)_{x+a+\oa_{\mu},x+\oa_{\mu}}(\oD_{\mu})_{x+\oa_{\mu},x}
+(\oD_{\mu})_{x+a+\oa_{\mu},x+a}(\D)_{x+a,x}.
\end{eqnarray}

Now several remarks are in order. 
Our current multiplet of $N=4$ SYM in three dimensions should contain
three components of gauge fields as well as three components of
scalar fields as the bosonic contents,
which can be interpreted as a dimensional reduction from 
six dimensional $N=1$ or four dimensional $N=2$ SYM.
It is then natural to require that the above bosonic gauge
link variables are to be defined 
in such a way to include the scalar contributions,
\begin{eqnarray}
(\U_{\pm \mu})_{x\pm n_{\mu},x} &\equiv&
(e^{\pm i(A_{\mu}\pm \phi^{(\mu)})})_{x\pm n_{\mu},x}, \label{exp}
\end{eqnarray}
where $A_{\mu}$ and $\phi^{(\mu)}(\mu =1,2,3)$ represent 
hermitian three dimensional gauge field and three components of scalar
fields, respectively.
Notice that the product of oppositely oriented bosonic gauge link variables
does not satisfy unitary nature, $\U_{+\mu}\U_{-\mu} \neq 1$,
and it leads to the contribution of scalar fields.
One could also see that, taking the na\"ive continuum limit,
through the expansion of %bosonic 
link variables,
\begin{eqnarray}
(\U_{\pm \mu})_{x\pm n_{\mu},x} 
&=& (1\pm i(A\pm \phi^{(\mu)})+\cdots)_{x\pm n_{\mu},x}\\
&\rightarrow& \mp(\partial_{\mu}-i(A_{\mu}\pm \phi^{(\mu)})), \label{cont_limit}
\end{eqnarray}
the gauge field $A_{\mu}$ actually transforms as a $SO(3)$ Lorentz vector
and an internal scalar while $\phi^{(\mu)}$ transforms as a Lorentz scalar 
and a $SO(3)$ internal vector. After the twisting, both of these bosonic components 
transform as $SO(3)$ vectors under the twisted Lorentz generator, $J_{\mu}'$,
which justifies the covariance of the expression (\ref{exp}).

The second remark is that the set of Leibniz rule conditions 
(\ref{leibniz1})-(\ref{leibniz3})
can now be interpreted as the gauge covariant conditions on the lattice
which restricts the orientation of bosonic gauge link variables, $\U_{\pm \mu}$,
on the r.h.s. of (\ref{constSYM1})-(\ref{constSYM3}).
These restrictions eventually pose strong constraints on possible complex nature
of gauge link variables,   
and as an inevitable consequence, 
%as long as our three-dimensional
%$N=4$ constraints 
%(\ref{constSYM1})-(\ref{constSYM4}) are concerned, 
one may not adopt the usual hermiticity conditions
on (\ref{constSYM1})-(\ref{constSYM4}),
although in the na\"ive continuum limit %it restores and 
one could actually obtains hermitian $N=4$ twisted SYM in three dimensions. 
It is important to notice that this issue is originated from the
peculiar structure of the three dimensional supercharges.
One could actually observe in the symmetric choice 
(Fig.\ref{3Dsymm_a})
that $a_{A}$ and $\overline{a}_{A}$ are located in the same orientations,
not in the opposite orientations as one might have expected to keep the hermiticity.
We recognize that this ``degenerate" lattice structure could be traced back to
the absence of chirality in three dimensions, namely the absence of $\gamma^{5}$ matrix. 
We keep these issues as a future investigation,
recognizing it would require yet further understanding of
supersymmetric lattice structure and lattice nature of chirality.
We then, in the following, proceed to perform an explicit construction of 
$N=4$ twisted SYM multiplet on three dimensional lattice.

After imposing the SYM constraints (\ref{constSYM1})-(\ref{constSYM4}),
Jacobi identities of three fermionic link variables give
\begin{eqnarray}
[\D_{\mu},\U_{+\nu}]_{x+a_{\mu}+n_{\nu},x}
 + [\D_{\nu},\U_{+\mu}]_{x+a_{\nu}+n_{\mu},x} &=& 0,  \\[4pt]
[\oD_{\mu},\U_{+\nu}]_{x+\oa_{\mu}+n_{\nu},x}
 + [\oD_{\nu},\U_{+\mu}]_{x+\oa_{\nu}+n_{\mu},x} &=& 0, \\[4pt]
[\D_{\mu},\U_{+\nu}]_{x+a_{\mu}+n_{\nu},x}
 +i\epsilon_{\mu\nu\rho}[\D,\U_{-\rho}]_{x+a-n_{\rho},x} &=& 0, \\[4pt]
[\oD_{\mu},\U_{+\nu}]_{x+\oa_{\mu}+n_{\nu},x}
 -i\epsilon_{\mu\nu\rho}[\oD,\U_{-\rho}]_{x+\oa-n_{\rho},x} &=& 0, \\[4pt]
\epsilon_{\mu\nu\lambda}[\D_{\rho},\U_{-\lambda}]_{x+a_{\rho}-n_{\lambda},x}
+\epsilon_{\rho\nu\lambda}[\D_{\mu},\U_{-\lambda}]_{x+a_{\mu}-n_{\lambda},x} &=& 0,\\[4pt]
\epsilon_{\mu\nu\lambda}[\oD_{\rho},\U_{-\lambda}]_{x+\oa_{\rho}-n_{\lambda},x}
+\epsilon_{\rho\nu\lambda}[\oD_{\mu},\U_{-\lambda}]_{x+\oa_{\mu}-n_{\lambda},x} &=& 0,\\[4pt]
[\D,\U_{+\mu}]_{x+a+n_{\mu},x} \ =\   
[\oD,\U_{+\mu}]_{x+\oa+n_{\mu},x} &=& 0,   
\end{eqnarray} 
where again all the commutators should be understood as link commutators.
In accordance with the above relations, one may define the following
non-vanishing fermionic link fields,
\begin{eqnarray}
[\D,\U_{-\rho}]_{x+a-n_{\rho},x} &\equiv& +(\olambda_{\rho})_{x-\oa_{\rho},x}, 
\label{fermion1} \\[2pt]
[\oD,\U_{-\rho}]_{x+\oa-n_{\rho},x} &\equiv& +(\lambda_{\rho})_{x-a_{\rho},x}, \\[2pt]
[\D_{\mu},\U_{+\nu}]_{x+a_{\mu}+n_{\nu},x} &=&
-i\epsilon_{\mu\nu\rho}(\olambda_{\rho})_{x-\oa_{\rho},x}, \\[2pt]
[\oD_{\mu},\U_{+\nu}]_{x+\oa_{\mu}+n_{\nu},x} &=&
+i\epsilon_{\mu\nu\rho}(\lambda_{\rho})_{x-a_{\rho},x}, \\[2pt]
[\D_{\mu},\U_{-\nu}]_{x+a_{\mu}-n_{\nu},x} 
&\equiv& +\delta_{\mu\nu}(\orho)_{x-\oa,x}, \\[2pt]
[\oD_{\mu},\U_{-\nu}]_{x+\oa_{\mu}-n_{\nu},x} 
&\equiv& +\delta_{\mu\nu}(\rho)_{x-a,x}, \label{fermion6}
\end{eqnarray}
where $(\olambda_{\mu},\lambda_{\mu},\orho,\rho)$
represent $N=4$ twisted fermions on three dimensional lattice.
%The shift properties of the twisted fermions are given in Table \ref{shift}. 
The Jacobi identities for two fermionic and one bosonic link variable
together with the relations (\ref{fermion1})-(\ref{fermion6})
give the following set of relations, 
\begin{eqnarray}
\{\D,\lambda_{\mu}\}_{x+a-a_{\mu},x}
 &=& +\frac{1}{2}\epsilon_{\mu\rho\sigma}
[\U_{+\rho},\U_{+\sigma}]_{x+n_{\rho}+n_{\sigma},x}, \\[0pt]
\{\oD,\olambda_{\mu}\}_{x+a-a_{\mu},x}
 &=& -\frac{1}{2}\epsilon_{\mu\rho\sigma}
[\U_{+\rho},\U_{+\sigma}]_{x+n_{\rho}+n_{\sigma},x}, \\[2pt]
\{\D_{\mu},\olambda_{\nu}\}_{x+a_{\mu}-\oa_{\nu},x} 
&=& -\delta_{\mu\nu}G_{x+a-\oa,x}, \\[2pt]
\{\oD_{\mu},\lambda_{\nu}\}_{x+\oa_{\mu}-a_{\nu},x} 
&=& -\delta_{\mu\nu}\oG_{x+\oa-a,x}, \\[2pt]
\{\D_{\mu},\lambda_{\nu}\}_{x+a_{\mu}-a_{\nu},x}
 &=& +i[\U_{+\mu},\U_{-\nu}]_{x+n_{\mu}-n_{\nu},x}
-\delta_{\mu\nu}(K_{x,x}+\frac{i}{2}[\U_{+\rho},\U_{-\rho}]_{x,x}), \qquad \\[0pt]
\{\oD_{\mu},\olambda_{\nu}\}_{x+\oa_{\mu}-\oa_{\nu},x}
 &=& +i[\U_{+\mu},\U_{-\nu}]_{x+n_{\mu}-n_{\nu},x}
+\delta_{\mu\nu}(K_{x,x}-\frac{i}{2}[\U_{+\rho},\U_{-\rho}]_{x,x}), \\[2pt]
\{\D,\olambda_{\mu}\}_{x+a-\oa_{\mu},x} 
&=& \{\oD,\lambda_{\mu}\}_{x+\oa-a_{\mu},x} \ =\ 0,
\end{eqnarray}
and
\begin{eqnarray}
\{\D_{\mu},\rho\}_{x+a_{\mu}-a,x} &=& +\frac{1}{2}\epsilon_{\mu\rho\sigma}
[\U_{-\rho},\U_{-\sigma}]_{x-n_{\rho}-n_{-\sigma},x}, \\[0pt]
\{\oD_{\mu},\orho\}_{x+\oa_{\mu}-\oa,x} &=& -\frac{1}{2}\epsilon_{\mu\rho\sigma}
[\U_{-\rho},\U_{-\sigma}]_{x-n_{\rho}-n_{-\sigma},x}, \\[0pt]
\{\D,\rho\}_{x,x} &=& -K_{x,x}+\frac{i}{2}[\U_{+\rho},\U_{-\rho}]_{x,x}, \\[0pt]
\{\oD,\orho\}_{x,x} &=& +K_{x,x}+\frac{i}{2}[\U_{+\rho},\U_{-\rho}]_{x,x}, \\[0pt]
\{\D,\orho\}_{x+a-\oa,x} &=& +G_{x+a-\oa,x}, \\[4pt]
\{\oD,\rho\}_{x+\oa-a,x} &=& +\oG_{x+\oa-a,x}, \\[2pt]
\{\D_{\mu},\orho\}_{x+a_{\mu}-\oa,x} &=& 
\{\oD_{\mu},\rho\}_{x+\oa_{\mu}-a,x} \ =\ 0, 
\end{eqnarray}
where $G$, $\oG$ and $K$ denote auxiliary fields defined on links 
$(x+a-\oa,x)$, $(x+\oa-a,x)$ and on a site, respectively.
All the shift properties of the component fields are summarized in 
Table \ref{shift}.

SUSY transformation of twisted $N=4$ lattice gauge multiplet
can be determined from the above Jacobi identity relations via
\begin{eqnarray}
(s_{A}\varphi)_{x+a_{A}+a_{\varphi},x} \ = \
(s_{A})\varphi_{x+a_{\varphi},x} &\equiv&
[\D_{A},\varphi\}_{x+a_{A}+a_{\varphi},x},
\label{SUSY-tr}
\end{eqnarray} 
where $(\varphi)_{x+a_{\varphi},x}$ denotes one of the component fields
$(\U_{\pm\mu},\rho,\olambda_{\mu},\lambda_{\mu},\orho,G,\oG,K)$.
The results are summarized in Table 2.
As a natural consequence of the constraints 
(\ref{constSYM1})-(\ref{constSYM4}),
one can see that the resulting $N=4\ D=3$ twisted SUSY algebra
for the component fields closes off-shell (modulo gauge transformations) on the lattice,
\begin{eqnarray}
\{s,\os_{\mu}\}(\varphi)_{x+a_{\varphi},x} 
&=& +i[\U_{+\mu},\varphi]_{x+n_{\mu}+a_{\varphi},x},\\[2pt]
\{s_{\mu},\os_{\nu}\}(\varphi)_{x+a_{\varphi},x}
&=& -\epsilon_{\mu\nu\rho}[\U_{-\rho},\varphi]_{x-n_{\rho}+a_{\varphi},x},\\[2pt]
\{\os, s_{\mu}\}(\varphi)_{x+a_{\varphi},x}
&=& +i[\U_{+\mu},\varphi]_{x+n_{\mu}+a_{\varphi},x}\\[2pt]
\{others\}(\varphi)_{x+a_{\varphi},x} &=& 0,
\end{eqnarray}
where again $\varphi$ denotes any component of the lattice multiplet
$(\U_{\pm\mu},\rho,\olambda_{\mu},\lambda_{\mu},\orho,G,\oG,K)$.

\begin{table}
\renewcommand{\arraystretch}{1.3}
\renewcommand{\tabcolsep}{5pt}
\hfil
\begin{tabular}{c|c|c|c|c|c}
\hline
& $\D$ & $\oD_{\mu}$ & $\D_{\mu}$ & $\oD$ & $\U_{\pm\mu}$  \\ \hline
shift & $a$ & $\oa_{\mu}$ & $a_{\mu}$ & $\oa$ & $\pm n_{\mu}$ 
\\ \hline
\end{tabular}
\hfil
\begin{tabular}{c|c|c|c|c|c|c|c}
\hline
& $\rho$ & $\olambda_{\mu}$ & $\lambda_{\mu}$ & $\orho$ & $G$ & $\oG$ & $K$ \\ \hline
shift & $-a$ & $-\oa_{\mu}$ & $-a_{\mu}$ & $-\oa$ & $a-\oa$ & $\oa-a$ & $0$
\\ \hline
\end{tabular}
\caption{Shifts carried by link variables and fields}
\label{shift}
\end{table}

\begin{table}[h]
\begin{center}
\renewcommand{\arraystretch}{1.25}
\renewcommand{\tabcolsep}{3pt}
\begin{tabular}{|c|c|c|c|c|}
\hline
& $s$ & $\os_{\mu}$ & $s_{\mu}$ & $\os$ \\ \hline
$\U_{+\nu}$ & $0$ & $+i\epsilon_{\mu\nu\rho}\lambda_{\rho}$
& $-i\epsilon_{\mu\nu\rho}\olambda_{\rho}$ & $0$ \\ 
$\U_{-\nu}$ & $+\olambda_{\nu}$ & $+\delta_{\mu\nu}\rho$ 
& $+\delta_{\mu\nu}\orho$ & $+\lambda_{\nu}$ \\ \hline
$\rho$ & $-K+\frac{i}{2}[\U_{+\rho},\U_{-\rho}]$ & $0$ 
& $+\frac{1}{2}\epsilon_{\mu\rho\sigma}[\U_{-\rho},\U_{-\sigma}]$ & $+\oG$ \\
$\olambda_{\nu}$ & $0$ 
& $+i[\U_{+\mu},\U_{-\nu}]$ 
& $-\delta_{\mu\nu}G$ 
& $-\frac{1}{2}\epsilon_{\nu\rho\sigma}[\U_{+\rho},\U_{+\sigma}]$
\\ %[-8pt]
& & $+\delta_{\mu\nu}(K-\frac{i}{2}[\U_{+\rho},\U_{-\rho}])$ & & \\
$\lambda_{\nu}$ 
& $+\frac{1}{2}\epsilon_{\nu\rho\sigma}[\U_{+\rho},\U_{+\sigma}]$ 
& $-\delta_{\mu\nu}\oG$ 
& $+i[\U_{+\mu},\U_{-\nu}]$ 
& $0$  
\\ %[-8pt]
& & & $-\delta_{\mu\nu}(K+\frac{i}{2}[\U_{+\rho},\U_{-\rho}])$ & \\
$\orho$ & $+G$ 
& $-\frac{1}{2}\epsilon_{\mu\rho\sigma}[\U_{-\rho},\U_{-\sigma}]$ 
& $0$ & $+K+\frac{i}{2}[\U_{+\rho},\U_{-\rho}]$ \\ \hline
$G$ & $0$ 
& $+\epsilon_{\mu\rho\sigma}[\U_{-\rho},\olambda_{\sigma}]$ 
& $0$
& $-i[\U_{+\rho},\olambda_{\rho}]$  
\\ %[-8pt]
& & $+i[\U_{+\mu},\orho]$ & &  \\
$\oG$ & $-i[\U_{+\rho},\lambda_{\rho}]$ & $0$ 
& $-\epsilon_{\mu\rho\sigma}[\U_{-\rho},\lambda_{\sigma}]$ & $0$ 
\\ %[-8pt]
& & & $+i[\U_{+\mu},\rho]$ &  \\
$K$ & $+\frac{i}{2}[\U_{+\rho},\olambda_{\rho}]$ 
& $+\frac{1}{2}\epsilon_{\mu\rho\sigma}[\U_{-\rho},\lambda_{\sigma}]$
& $+\frac{1}{2}\epsilon_{\mu\rho\sigma}[\U_{-\rho},\olambda_{\sigma}]$
& $-\frac{i}{2}[\U_{+\rho},\lambda_{\rho}]$ 
\\ %[-6pt]
& & $-\frac{i}{2}[\U_{+\mu},\rho]$ & $+\frac{i}{2}[\U_{+\mu},\orho]$ & \\ 
\hline 
\end{tabular}
\label{translat3DSYM}
\caption{SUSY trans. laws for twisted $N=4\ D=3$ lattice SYM multiplet
$(\U_{\pm\mu},\rho,\olambda_{\mu},\lambda_{\mu},\orho,G,\oG,K)$}
\end{center}
\end{table}

One of the important properties of the above multiplet and SUSY
transformations is that each $\U_{\pm\mu}$ satisfies 
``chiral" or ``anti-chiral" condition. 
See Ref.\cite{DKKN1} for the ``chiral" conditions.  
For example, $\U_{+3}$ and $\U_{-3}$ satisfy,  
\begin{eqnarray}
s\ \U_{+3} \ =\ \os_{3}\ \U_{+3} \ =\ s_{3}\ \U_{+3}\ =\
\os\ \U_{+3} &=& 0, \label{ccD3_1} \\[2pt]
s_1\ \U_{-3} \ =\ s_{2}\ \U_{-3} \ =\ \os_{1}\ \U_{-3}\ =\
\os_{2}\ \U_{-3} &=& 0, \label{ccD3_2}
\end{eqnarray}
and  similar relations hold  
for $\U_{\pm 1}$ and $\U_{\pm 2}$. 
One could thus observe that the twisted $N=4\ D=3$ lattice SUSY invariant 
action can be manifestly constructed by, for example,
successive operations of $\os_{1}\os_{2}s_{1}s_{2}$
on $\U_{+3}\U_{+3}$
or $s\os_{3}\os s_{3}$ on $\U_{-3}\U_{-3}$.
These two combinations turn out to be equivalent
each other and give,
\begin{eqnarray}
S&=& +\sum_{x}\frac{1}{2}\os_{1}\os_{2}s_{1}s_{2}
\ \mathrm{tr}\ \U_{+3}\ \U_{+3} 
\ =\ -\sum_{x}\frac{1}{2}s \os_{3}\os s_{3}
\ \mathrm{tr}\ \U_{-3}\ \U_{-3}\\ 
\nonumber
&=& \sum_{x}\mathrm{tr}\biggl[\
\frac{1}{4}[\U_{+\mu},\U_{-\mu}]_{x,x}[\U_{+\nu},\U_{-\nu}]_{x,x} 
+K^{2}_{x,x} \nonumber \\ 
&&-\frac{1}{2}%\epsilon_{\mu\rho\sigma}\epsilon_{\mu\alpha\beta}
[\U_{+\mu},\U_{+\nu}]_{x,x-n_{\mu}-n_{\nu}}
[\U_{-\mu},\U_{-\nu}]_{x-n_{\mu}-n_{\nu},x} 
+G_{x,x+\oa-a}\oG_{x+\oa-a,x}\nonumber \nonumber \\[2pt]
&&+i(\olambda_{\mu})_{x,x+\oa_{\mu}}[\U_{+\mu},\rho]_{x+\oa_{\mu},x} 
+ i(\lambda_{\mu})_{x,x+a_{\mu}}
[\U_{+\mu},\orho]_{x+a_{\mu},x} 
+\epsilon_{\mu\nu\rho}(\lambda_{\mu})_{x,x+a_{\mu}}
[\U_{-\nu},\olambda_{\rho}]_{x+a_{\mu},x}
\biggr], \nonumber \\
\label{N=4D=3action}
\end{eqnarray}
where the summation of $x$ should cover integer sites as well as half-integer sites 
if one takes the symmetric choice of $a_{A}$ (\ref{3Dsymm1})-(\ref{3Dsymm4}),
\begin{eqnarray}
\sum_{x,\ symm\ a_{A}} &=& \sum_{(m_{1},m_{2},m_{3})} 
+ \sum_{(m_{1}+\frac{1}{2},m_{2}+\frac{1}{2},m_{3}+\frac{1}{2})},\ \ 
(m_{1},m_{2},m_{3} :\mathrm{integers}),
\end{eqnarray}
while for the asymmetric choice of $a_{A}$ (\ref{3Dasymm1})-(\ref{3Dasymm4}),
 it needs to cover only the integer sites,
\begin{eqnarray}
\sum_{x,\ asym\ a_{A}} &=& \sum_{(m_{1},m_{2},m_{3})},\ \ 
(m_{1},m_{2},m_{3} : \mathrm{integers}).
\end{eqnarray} 
Due to this summation property, the order in the product of 
supercharges is shown to be irrelevant up to
total difference terms.
Notice that the exact form w.r.t. all the supercharges
and the nilpotency of each supercharge manifestly ensure
the twisted $N=4$ SUSY invariance of the action. 
%\footnote{Although the invariance of the action (\ref{N=4D=3action})
%is algebraically ensured by the nilpotency of the supercharges,
%we recognize that the SUSY variations of component fields
%should be supplemented by lattice counterparts of
%Grassmann parameters 
%such that the variations maintain
%the gauge covariance,
%as is pointed out in \cite{BC}.}
It is also important to note that each term in the action
forms closed loop, which ensures manifest gauge invariance of the action.
This property is originated from the 
vanishing sum of the shifts associated with the action,
\begin{eqnarray}
\oa_{1}+\oa_{2}+a_{1}+a_{2}+n_{3}+n_{3}\ =\ 
a+\oa_{3}+\oa+a_{3} -n_{3}-n_{3} &=& 0,
\end{eqnarray} 
which holds for any particular choice of $a_{A}$.
The gauge invariance is thus maintained
regardless of any particular choice of $a_{A}$.

The invariance of the action under SUSY transformations 
originates essentially from the fact that it is exact with respect to all 
eight nilpotent supersymmetry charges as we can see from 
(\ref{N=4D=3action}). 
However the modified Leibniz rule appears to introduce some ambiguity in
the supersymmetry variation of products of fields. This criticism was 
formulated in ~\cite{Bruckmann} within the framework of an $N=2$ 
supersymmetric quantum mechanics. It was argued that the supersymmetry 
transformation of a product of two component fields depends on the order 
in which the product is written even if the fields themselves commute. 
Hence the whole approach was claimed in ~\cite{Bruckmann} to be inconsistent. 
%The answer to this apparent inconsistency has been given 
A possible answer to this criticism has been given
within the same model 
in the Lattice 2007 Proceedings ~\cite{ADKS} (a more extensive 
paper \cite{ADKS} will follow). It is shown there that no ambiguity 
whatsoever is present if the modified Leibniz rule is applied to superfields 
products when performing a SUSY variation. At the level of component fields 
this means that when applying the modified Leibniz rule the order of the 
fields in the different terms of the action must reflect the original order 
of the superfield product, even if the fields themselves commute. 
In fact, due to the slightly non local nature of superfields on the lattice 
as defined in  ~\cite{DKKN1}, superfield products are intrinsically 
non-commutative even if the product, for instance, of their first components 
is commutative. We explain more details on this controversial issue in 
Appendix. The non-commutativity of the superfields product and the 
modified Leibniz rule can be understood in terms of non-commutative geometry
as a result of a special case of Moyal product defined on the lattice. 
The details are given in ~\cite{ADKS}.

The criticism of ref. ~\cite{Bruckmann} was extended in ~\cite{BC} to the
case of gauge theories and to the link approach formulated in ~\cite{DKKN2},
which is more relevant to the present paper.
The "inconsistency" claimed in ~\cite{BC} is related to the link 
nature of supercharge $s_A$ and supercovariant derivative $\D_A$. 
A SUSY transformation $s_A$ on the action generates a link hole 
$(x+a_A,x)$ since all the terms in the action have a vanishing shift 
and thus are composed of closed loops. At a first look a naive super 
charge operation to the action leads to gauge variant terms since 
the terms have the link holes. We claim that we need to introduce 
covariantly constant fermionic parameter $\eta_B$ which anti-commutes 
with all supercovariant derivatives in the shifted anti-commutator sense, 
\begin{eqnarray}
\{\D_A,\eta_B\}_{x+a_A-a_B,x} %\nonumber \\
&=& (\D_A)_{x+a_A-a_B,x-a_B}(\eta_B)_{x-a_B,x} + 
(\eta_B)_{x+a_A-a_B,x+a_A}(\D_A)_{x+a_A,x} \nonumber \\
 &=& 0,
\label{fermionic-P}
\end{eqnarray} 
%\begin{eqnarray}
%\lefteqn{\{\D_A, \eta_B\}_{x+a_A-a_B,x}}\quad \nonumber \\
%&=& (\D_A)_{x+a_A-a_B,x-a_B}(\eta_B)_{x-a_B,x} +
%(\eta_B)_{x+a_A-a_B,x+a_A}(\D_A)_{x+a_A,x} \nonumber \\
% &=& 0,
%\label{fermionic-P}
%\end{eqnarray}
%\begin{eqnarray}
%\{\D_A&,&\eta_B\}_{x+a_A-a_B,x} \nonumber \\
%~&=&~ (\D_A)_{x+a_A-a_B,x-a_B}(\eta_B)_{x-a_B,x} + 
%(\eta_B)_{x+a_A-a_B,x+a_A}(\D_A)_{x+a_A,x} \nonumber \\
% &=& 0,
%\label{fermionic-P}
%\end{eqnarray} 
where $\eta_B$ has a shift $-a_B$ and thus can fill up the link holes 
to generate gauge invariant terms. 
We define the gauge transformation of the superparameter, 
\begin{eqnarray}
(\eta_A)_{x-a_A,x} \rightarrow G_{x-a_A}(\eta_A)_{x-a_A,x}G_x^{-1}.
\label{gauge-tr-FP}
\end{eqnarray} 
We can then prove the exact SUSY invariance of the action by applying a 
shiftless combination of SUSY transformation $\eta_A s_A$ (no sum) to the 
action. 
SUSY transformation of component fields with fermionic parameter is given 
by 
\begin{eqnarray}
(\eta_A s_{A}\varphi)_{x+a_{\varphi},x} \ = \
(\eta_A)_{x+a_{\varphi},x+a_{\varphi}+a_A} 
(s_{A}\varphi)_{x+a_{\varphi}+a_A,x},
\label{SUSY-tr-CF}
\end{eqnarray} 
where the SUSY transformation of $(s_{A}\varphi)_{x+a_{\varphi}+a_A,x}$ 
is defined by (\ref{SUSY-tr}) and is given in the Table 2. 

\begin{figure}[h]
\begin{center}
\includegraphics[width=100mm]{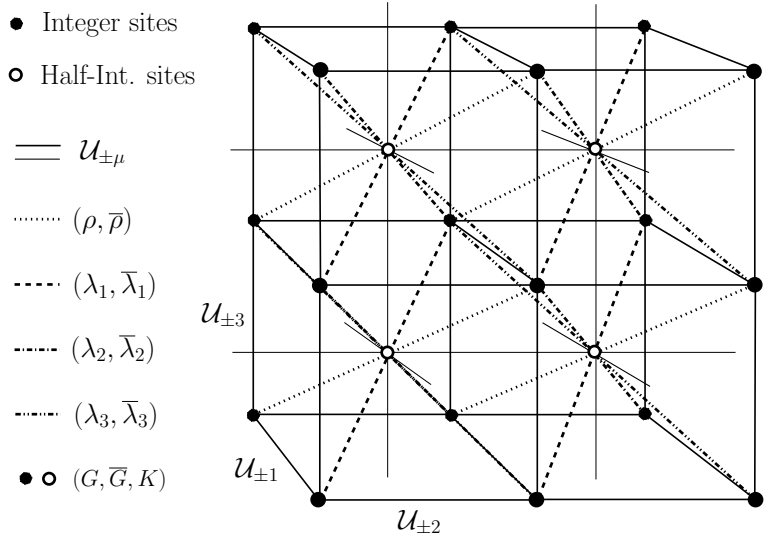}
\caption{All the configurations in $N=4\ D=3$ twisted SYM action for symmetric $a_{A}$}
\label{3Dallconfig}
\end{center}
\end{figure}

\begin{figure}[h]
\begin{center}
\includegraphics[width=40mm]{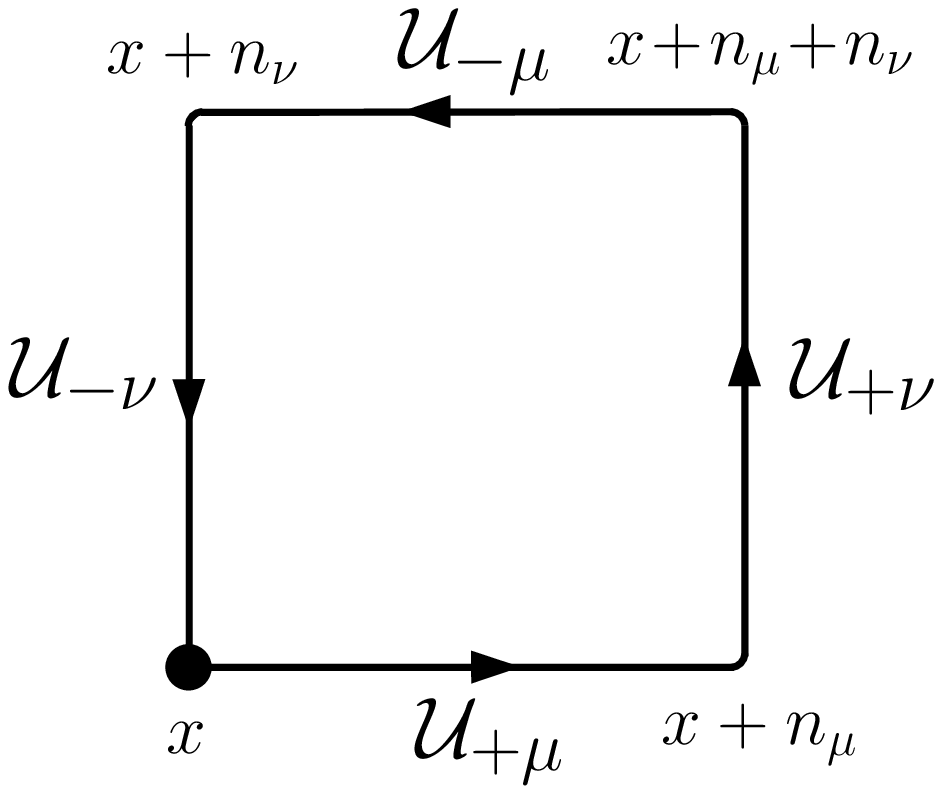}
\includegraphics[width=40mm]{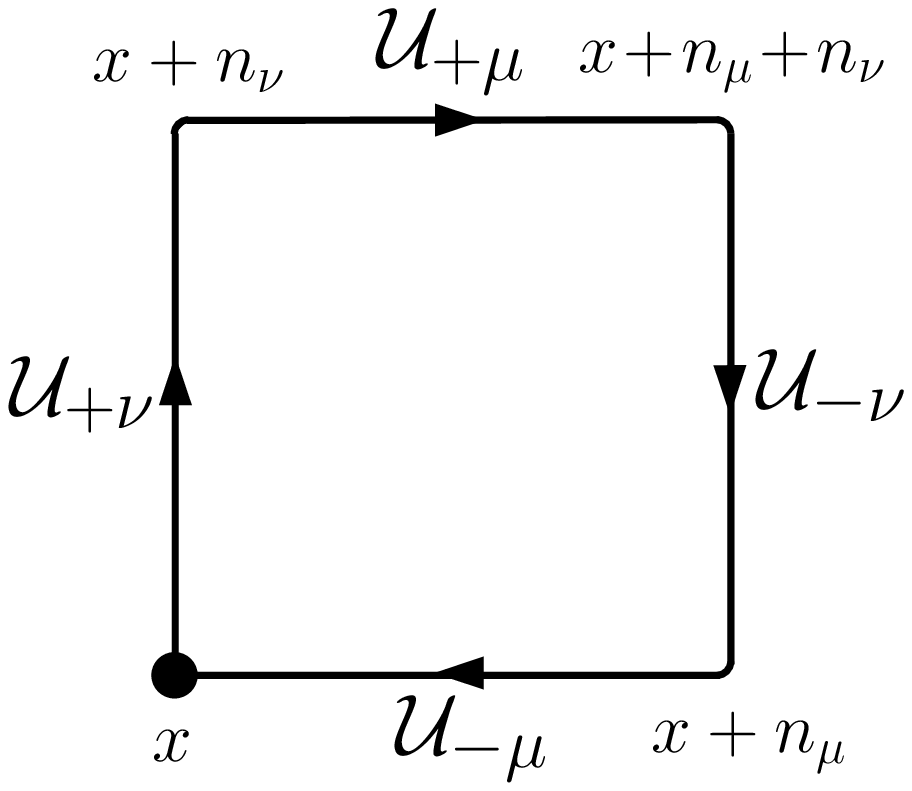}
\caption{Plaquettes}
\label{3DSYMPlaquettes}
\end{center}
\end{figure}

\begin{figure}
\begin{center}
\includegraphics[width=35mm]{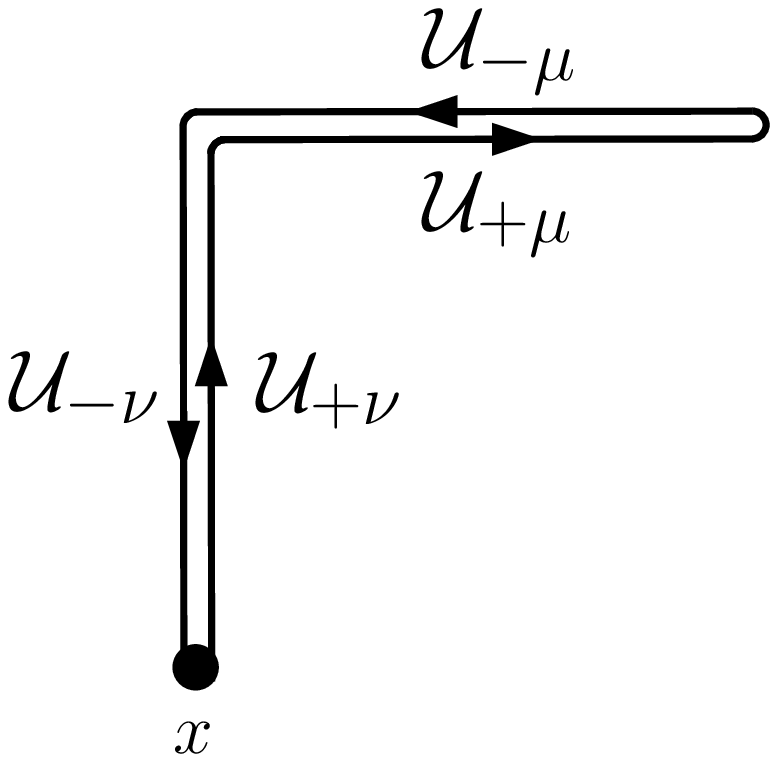}
\includegraphics[width=35mm]{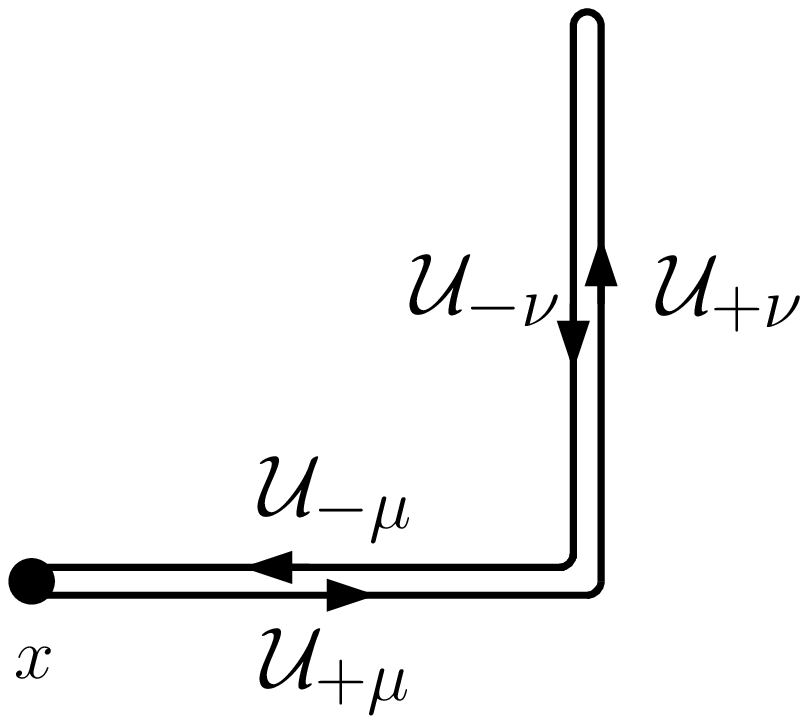}
\includegraphics[width=35mm]{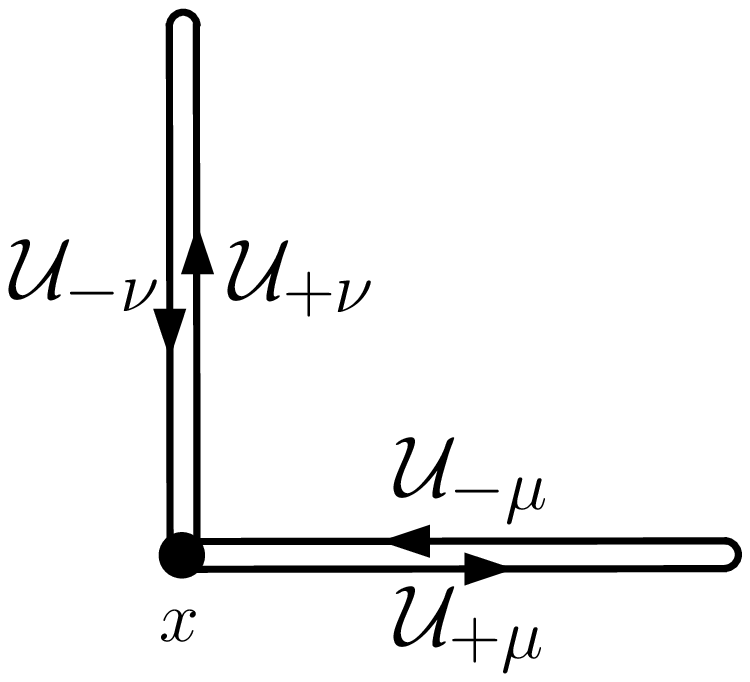}
\caption{Zero area loops}
\label{3DSYMzero_area}
\end{center}
\end{figure}

\begin{figure}[h]
\begin{center}
\includegraphics[width=40mm]{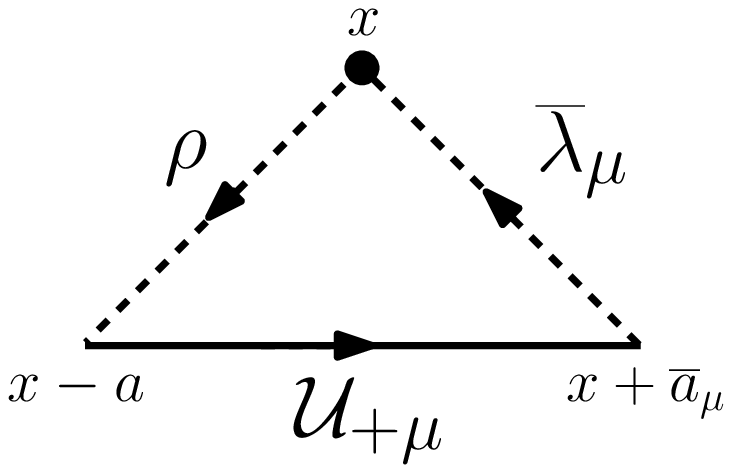}
\hspace{10pt}
\includegraphics[width=40mm]{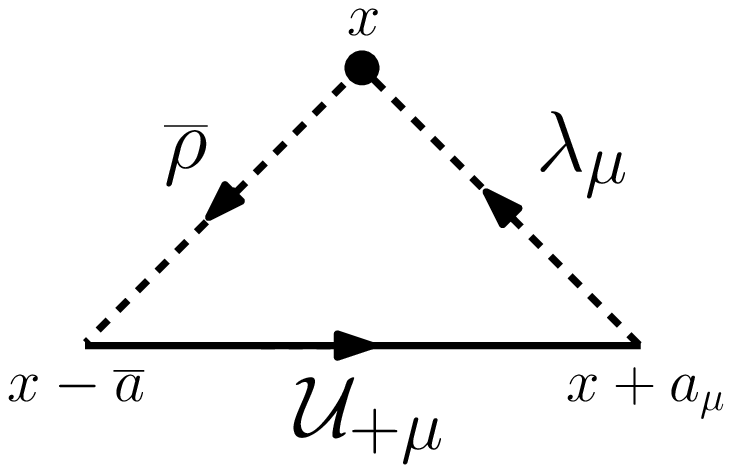}
\hspace{10pt}
\includegraphics[width=40mm]{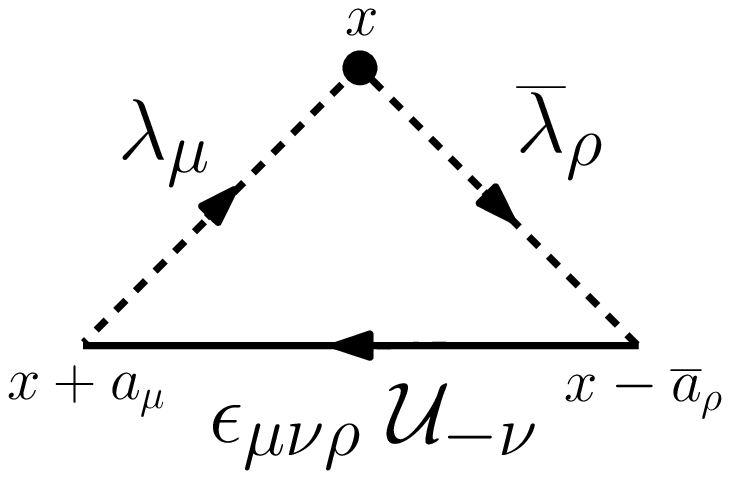}
\caption{Fermion loops in $N=4\ D=3$ twisted SYM action}
\label{3DSYMfermionloops}
\end{center}
\end{figure}

Fig.\ref{3Dallconfig} depicts all the field configurations
in $N=4\ D=3$ twisted SYM action (\ref{N=4D=3action})
in the case of symmetric choice of $a_{A}$, (\ref{3Dsymm1})-(\ref{3Dsymm4}),
where the bosonic gauge link variables $\U_{\pm\mu}$ are
located on solid links, while fermionic link components 
$(\rho,\olambda_{\mu},\lambda_{\mu},\orho)$
are located on diagonal links.
Notice that in the case of symmetric choice of $a_{A}$,
only the auxiliary fields $(G,\overline{G},K)$ are located on sites.
The bosonic part of the action consists of 
usual plaquette terms (Fig.\ref{3DSYMPlaquettes})  
as well as zero-area loops which represent the contributions
of scalar fields $\phi^{(\mu)}$ (Fig.\ref{3DSYMzero_area})
which, as mentioned above, are originated from the property 
$\U_{+\mu}\U_{-\mu}\neq 1$.
Fermion terms in the action (\ref{N=4D=3action})
consist of closed triangle loops (Fig. \ref{3DSYMfermionloops}). 


The na\"ive continuum limit of the action (\ref{N=4D=3action})
can be taken through the expansion of gauge link variables 
(\ref{cont_limit}). After using trace properties,
one obtain the following continuum action,
\begin{eqnarray}
S\rightarrow S_{cont}
&=& \int d^{3}x\ \mathrm{tr}\ \biggl[
\frac{1}{2}%[\Dc_{\mu},\Dc_{\nu}][\Dc_{\mu},\Dc_{\nu}]  
F_{\mu\nu}F_{\mu\nu}
+K^{2} + G\oG \nonumber \\
&& -[\Dc_{\mu},\phi^{(\nu)}][\Dc_{\mu},\phi^{(\nu)}]
-\frac{1}{2}[\phi^{(\mu)},\phi^{(\nu)}][\phi^{(\mu)},\phi^{(\nu)}] \nonumber \\[6pt]
&&-i\olambda_{\mu}[\Dc_{\mu},\rho] - i\lambda_{\mu}[\Dc_{\mu},\orho]
+\epsilon_{\mu\nu\rho}\lambda_{\mu}[\Dc_{\nu},\olambda_{\rho}]\nonumber \\[4pt]
&& -\olambda_{\mu}[\phi^{(\mu)},\rho] - \lambda_{\mu}[\phi^{(\mu)},\orho]
+i\epsilon_{\mu\nu\rho}\lambda_{\mu}[\phi^{(\nu)},\olambda_{\rho}]
\biggr], \label{3DSYMaction_cont2}
\end{eqnarray}
where $F_{\mu\nu}\equiv i[\Dc_{\mu},\Dc_{\nu}]$ represents the field strength
with $\Dc_{\mu} \equiv \partial_{\mu} -iA_{\mu}$, while $\phi^{(\mu)}(\mu=1,2,3)$
denote three independent hermitian scalar fields
in the twisted $N=4\ D=3$ SYM multiplet in the continuum spacetime. 
One could see that the kinetic term and potential term as well as Yukawa
coupling terms for scalar fields naturally come up from the
contributions of zero-area loops in the lattice action. 
The above action 
(\ref{3DSYMaction_cont2}) have complete agreement with
continuum construction of $N=4$ twisted SYM in three dimensions.

\section{Discussions} 

A fully exact SUSY invariant formulation of 
twisted $N=4$ SYM action on three dimensional lattice 
is presented. 
Algebraic relations of Jacobi identities are geometrically 
realized on the simplicial lattice with the help of shift 
relations of component fields. 
The three dimensional lattice structure embedding the twisted $N=4$ SUSY
naturally appears from the intrinsic relation
between twisted fermions and Dirac-K\"ahler fermions.
Twisted $N=4$ SUSY invariance is a natural consequence of the exact 
form of the action with respect to all the twisted supercharges 
up to surface terms which naturally vanish due to a trace property 
on the lattice. 

Possible answers to the critiques on the formulation of link 
approach are given. 
It is pointed out that there is a proper ordering of a product 
of component fields which leads correct lattice SUSY transformation. 
We needed to introduce superparameters which anti-commute with all the 
supercovariant derivatives. 
It would be important to find an explicit representation of the super 
parameters. We further have to accept that a structure behind the nature
of component fields which carry a shift and satisfy the relation 
(\ref{field-NC}) still remains to be better clarified.
We consider that the lattice SUSY transformation can be defined only 
semilocally due to the next neighboring ambiguity of difference operation 
and thus gives influence on the ordering of component fields. 
Superfield may be able to take care of this semilocal nature of SUSY 
transformation faithfully\cite{ADKS}.  

In this paper we have not addressed the issue of hermiticity of 
the formulation. 
Although we haven't yet reached to a complete understanding of
hermitian property on odd dimensional lattice, it is possible 
to understand hermitian property and Majorana nature of fermion 
in two dimensional formulation\cite{DKKN3}.
We recognize that hermitian property of lattice SYM should also be
understood from this perspective together with a better geometrical 
understandings of chirality on the lattice.
It should also be mentioned that a dimensional reduction of 
three dimensional $N=4$ twisted SYM
could give us a formulation of $N=4$ twisted SYM 
on two dimensional lattice, which corresponds to a double charged
system of $N=D=2$ twisted SYM\cite{DKKN3}.
It is also important to proceed to perform a possible lattice formulation of 
$N=D=4$  Dirac-K\"ahler twisted SYM 
which should be carried out basically in the same manner as
presented here.     
The results of these analyses will be given elsewhere.

\subsection*{Acknowledgments}

We would like to thank to J.~Kato, A.~Miyake and J.~Saito for useful discussions.
This work is supported in part by Japanese Ministry of Education,
Science, Sports and Culture under the grant number 18540245 and  
also by INFN research funds.
I.K. is supported by the Special Postdoctoral Researchers Program at RIKEN.
K.N. is supported by Department of Energy US Government, 
Grant No. FG02-91ER 40661.

%\appendix
%\section{Appendix}

%\begin{center}
%{\Large \textbf{Appendix}\\[0pt]}
%\end{center}
\renewcommand{\theequation}{A-\arabic{equation}}
\setcounter{equation}{0}
\section*{APPENDIX}

In this appendix we briefly discuss the argument given in \cite{Bruckmann} 
which according to the authors leads to an inconsistency in our lattice SUSY  
formulation with modified Leibniz rule and we show that there is no inconsistency at all. 
In this Appendix we consider, as in  \cite{Bruckmann}, only the case of the non-gauge
version of lattice SUSY formulation. 
The lattice formulation of supersymmetric gauge theories, which is  more relevant 
for the present paper and was criticized in \cite{BC}, has been briefly discussed 
in Section 3.   

Let us consider two superfields $\Phi_i(x,\theta_A)$ ($i=1,2$) whose expansion  
into component fields is given by \cite{DKKN1}: 
\begin{align}    
\Phi_1(x,\theta_A)&= \phi_1(x) + \theta_A \psi_1^A(x)+\cdots \nonumber\\     
\Phi_2(x,\theta_A)&= \phi_2(x) + \theta_A \psi_2^A(x)+\cdots ,     
\label{superfield-A}. 
\end{align}  
Here we assume that the superfields 
$\Phi_1(x)$ and $\Phi_2(x)$ do not  carry any shift while, 
according to \cite{DKKN1}, $\theta_A$ carry  a shift opposite to that of the 
super charge $Q_A$ and thus $\psi_A$ carry  the same shift $a_A$ as $Q_A$.  
%As we can see the superfields in general have a slightly nonlocal feature 
%due to the shift property of the super coordinates.  
The following notation will be used: given a superfield $\Phi(x,\theta_A)$ 
we denote with $\Phi|_0 (x)$ its first component, and by $\Phi|_A(x)$ 
the component corresponding to the coefficient of $\theta_A$ in the expansion. 
So for instance in (\ref{superfield-A}) we have: 
\begin{equation} 
\Phi_i|_0(x) = \phi_i(x),~~~~~~~~~~~~~~~\Phi_i|_A(x) = \psi_i^A(x). 
\label{comp} 
\end{equation} 
We define a super symmetry transformation $\delta_A=\eta_AQ_A$ (no sum) where  
$\eta_A$ is a supersymmetry parameter which carry the opposite shift of $Q_A$ 
so that the $\delta_A$ operation is shiftless. 
The SUSY transformation  of the component fields can then be easily obtained, 
and we shall focus here on the transformation properties of the first component 
of a superfield (the argument can be extended to higher components)which reads: 
\begin{equation} 
\delta_A \Phi_i|_0(x)\equiv (\delta_A \Phi_i)|_0(x)  
= \eta_A \Phi_i|_A(x), 
\label{susytr} 
\end{equation} 
that is, using (\ref{comp}) 
\begin{equation}  
\delta_A \phi_i(x)=\eta_A\psi_i^A(x). 
\label{SUSYtr-A} 
\end{equation}  
According to \cite{Bruckmann} an inconsistency occurs when the SUSY variation 
of the product $\phi_1(x)\phi_2(x)$ is considered. 
Indeed if the variation  $\delta_A(\phi_1\phi_2)(x)$ is calculated using 
the modified Leibniz rule of eq. (\ref{ops}) the result will depend on the order 
of the two fields within the variation symbol in spite of the fact that they commute: 
of two super fields with a different order, we obtain  
\begin{align}  
\delta_A(\phi_1(x)\phi_2(x))&=\eta_A(\psi_1^A(x)\phi_2(x) +  
\phi_1(x+a_A)\psi_2^A(x)),  \nonumber \\  
\delta_A(\phi_2(x)\phi_1(x))&=\eta_A(\psi_2^A(x)\phi_1(x) +  
\phi_2(x+a_A)\psi_1^A(x)),  \label{A3} 
\end{align} 
where we have used the following relation:  
\begin{equation} 
\phi_i(x)\eta_A=\eta_A\phi_i(x+a_A). 
\label{noncommutativity-of-parameter} 
\end{equation}  
As a result the authors of \cite{Bruckmann} claim that 
"there is an ambiguity in showing supersymmetry invariance of 
lattice actions" as a lattice action may be, or may be not, 
supersymmetric invariant depending on the order in which certain products 
of commuting component fields are written. 
In order to show that no ambiguity is really present we consider the product 
of the two superfields $\Phi_1$ and $\Phi_2$ and remark due to the noncommutativity 
between $\theta_A$ and the component fields as in  
(\ref{noncommutativity-of-parameter}), we have: 
\begin{equation} 
\Phi_1\Phi_2\neq \Phi_2 \Phi_1.  
\label{NC-superfields} 
\end{equation} 
The non-commutativity of the superfield product however does not apply to its 
lowest component (where no $\theta_A$ is involved) so that we have: 
\begin{equation} 
\Phi_1\Phi_2|_0(x) = \phi_1(x)\phi_2(x) 
= \phi_2(x)\phi_1(x) = \Phi_2\Phi_1 |_0(x), 
\label{lcomp} 
\end{equation} 
whereas we have 
\begin{equation} \Phi_1\Phi_2|_A(x) \neq \Phi_2\Phi_1|_A(x). 
\label{Acomp} 
\end{equation} 
This is really the crucial point of the whole issue: 
$\phi_1(x)\phi_2(x) = \phi_2(x)\phi_1(x)$ can be seen as the first component 
of two (slightly) different superfields, namely $\Phi_1\Phi_2$ and $\Phi_2 \Phi_1$.  
This does not happen in the continuum, where superfields commute and the first 
component identifies the superfield completely. 
We can now write the SUSY transformations (\ref{A3}), 
using the notation of  (\ref{susytr}):  
\begin{align}  
\delta_A(\Phi_1\Phi_2|_0(x))&=\eta_A (\Phi_1\Phi_2)|_A(x),  \nonumber \\  
\delta_A(\Phi_2\Phi_1|_0(x))&=\eta_A(\Phi_2 \Phi_1)|_A(x). \label{A3bis} 
\end{align} 
There is no ambiguity in eq. (\ref{A3bis}) although the arguments of $\delta_A$ 
happen to coincide due to (\ref{lcomp}). 
In order to give an unambiguous meaning to (\ref{A3}) we have to 
%convene 
agree that, 
although $\phi_1(x)$ and $\phi_2(x)$ commute and the numerical value of 
$ \phi_1(x)\phi_2(x)$ and of $ \phi_2(x)\phi_1(x)$ coincide, 
$ \phi_1(x)\phi_2(x)$ (resp. $ \phi_2(x)\phi_1(x)$) should be read as $\Phi_1\Phi_2|_0(x)$ 
(resp. $\Phi_2\Phi_1|_0(x)$) whenever used within a SUSY variation symbol $\delta_A$ 
in order to apply correctly the modified Leibniz rule. 
It should be noted that $\theta_A$ expansion of
$\Phi_1\Phi_2$ and $\Phi_2\Phi_1$ have different expressions as a product
of component fields except for the lowest component.
This means that the order in which the component fields appear in the different terms 
of a Lagrangian is relevant if the correct SUSY transformation are to be reproduced 
via the modified Leibniz rule. 
We claim that one particular ordering is the correct one: 
the one that reflects in each term of the lagrangian the original order of 
the superfield product. 
Needless to say that if one uses the superfield formalism consistently the problem 
does not arise at all. 
It would be tempting then to require a complete commutativity of the superfields 
on the lattice. 
This could be achieved at the expense of introducing some non-commutativity between 
shifted component fields, namely:  
\begin{equation} 
\phi_A(x+a_B)\phi_B(x)~~=~~ (-1)^{|\phi_A||\phi_B|}\phi_B(x+a_A)\phi_A(x),  
\label{field-NC} 
\end{equation}  
where the fields $\phi_A$ and $\phi_B$ carry a shift $a_A$ and $a_B$, respectively. 
If (\ref{field-NC}) was satisfied the two expressions at the r.h.s. of (\ref{A3}) 
would coincide and any field order could be adopted in the lagrangian, starting 
from the "correct" one, provided the coordinates were shifted according to 
(\ref{field-NC}). However, unless a concrete representation of $\phi_A(x)$ is 
found that satisfy (\ref{field-NC}), the condition above remains purely formal 
and difficult to use, say, within a functional integral.

%%%%%%%%%%%%%%%%%%%%%%%%%%%%%%%%%%%%%%%%%%%%%%%%%%%%%%%%%%%%%%%%%%%%%%%%%%%%
%%%%%
%%%%%%%%%%%%%%%%%%%%%%%%%%%%%%%%%%%%%%%%%%%%%%%%%%%%%%%%%%%%%%%%%%%%%%%%%%%%
%%%%%

\end{document}